# On the Child–Langmuir law in one, two, and three dimensions

Y. Y. Lau,[1,a)] Dion Li,[1] and David P. Chernin[2]

## AFFILIATIONS

[1]Department of Nuclear Engineering and Radiological Sciences, University of Michigan, Ann Arbor, Michigan 48109, USA
[2]Leidos, Inc., Reston, Virginia 20190, USA

[a)]Author to whom correspondence should be addressed: yylau@umich.edu

## ABSTRACT

We consider the limiting current from an emitting patch whose size is much smaller than the anode–cathode spacing. The limiting current is formulated in terms of an integral equation. It is solved iteratively, first to numerically recover the classical one-dimensional Child–Langmuir law, including Jaffe's extension to a constant, nonzero electron emission velocity. We extend to two-dimensions in which electron emission is restricted to an infinitely long stripe with infinitesimally narrow stripe width so that the emitted electrons form an electron sheet. We next extend to three-dimensions in which electron emission is restricted to a square tile (or a circular patch) with an infinitesimally small tile size (or patch radius) so that the emitted electrons form a needlelike line charge. Surprisingly, for the electron needle problem, we only find the null solution for the total line charge current, regardless of the assumed initial electron velocity. For the electron sheet problem, we also find only the null solution for the total sheet current if the electron emission velocity is assumed to be zero, and the total maximum sheet current becomes a finite, nonzero value if the electron emission velocity is assumed to be nonzero. These seemingly paradoxical results are shown to be consistent with the earlier works of the Child–Langmuir law of higher dimensions. They are also consistent with, or perhaps even anticipated by, the more recent theories and simulations on thermionic cathodes that used realistic work function distributions to account for patchy, non-uniform electron emission. The mathematical subtleties are discussed.

## I. INTRODUCTION

Non-uniform electron emission from a cathode surface is notoriously difficult to characterize.[1] One key question, which remains unanswered, is the maximum allowed average current density and its relation to the classical Child–Langmuir law (CLL),[2,3] which gives the maximum uniform steady state current density that can be transported across a planar diode of gap voltage $V$ and gap separation $d$ [Fig. 1(a)],

$$J_{CL} = \frac{4\sqrt{2}}{9}\epsilon_0\sqrt{\frac{e}{m}}\frac{V^{3/2}}{d^2}, \tag{1.1}$$

where $e$ and $m$ are, respectively, the electron charge and mass and $\epsilon_0$ is the free space permittivity. Equation (1.1) is a constraint imposed by the Poisson equation and the continuity equation in a one-dimensional (1D), planar, nonrelativistic diode. It is independent of the cathode's material properties. However, emission from a cathode is generally non-uniform and is highly dependent on the cathode temperature, material properties, emission processes, and surface roughness. Strong electron emission from a localized spot, whose size is small compared with the anode–cathode (AK) gap spacing $d$, is also a common occurrence, though rarely understood, or analyzed, in its relation to the CLL. This paper examines this issue.

The literature on non-uniform cathode emission is extensive.[1,4–29] Extending the classical 1D CLL to 2D in order to understand some aspects of non-uniform emission, Luginsland *et al.*[8] performed particle-in-cell simulations in which a uniform emission current density was assumed to occur over a finite stripe of width $W$ in a planar gap of separation $d$ [Fig. 1(b)]. They arrived at the following 2D scaling law, synthesized from their simulation data,

$$\frac{J(2)}{J_{CL}} \cong 1 + \frac{d}{\pi W}, \quad 0 < d/W < 10. \tag{1.2}$$

Lau[9] later analytically derived the scaling law (1.2) under the assumption $d/W \ll 1$. They found that this scaling law fit the numerical data to within a few percent and that it is virtually independent of an external magnetic field (ranging from 0 to 100 T) imposed longitudinally along the electron flow direction. Assuming uniform emission of electrons over a circular patch of radius $R$ with $d/R \ll 1$, a similar 2D scaling law was derived,[9]

$$\frac{J(2)}{J_{CL}} \cong 1 + \frac{d}{4R}, \quad 0 < d/R < 2, \tag{1.3}$$

which also fit the numerical data to within a few percent.

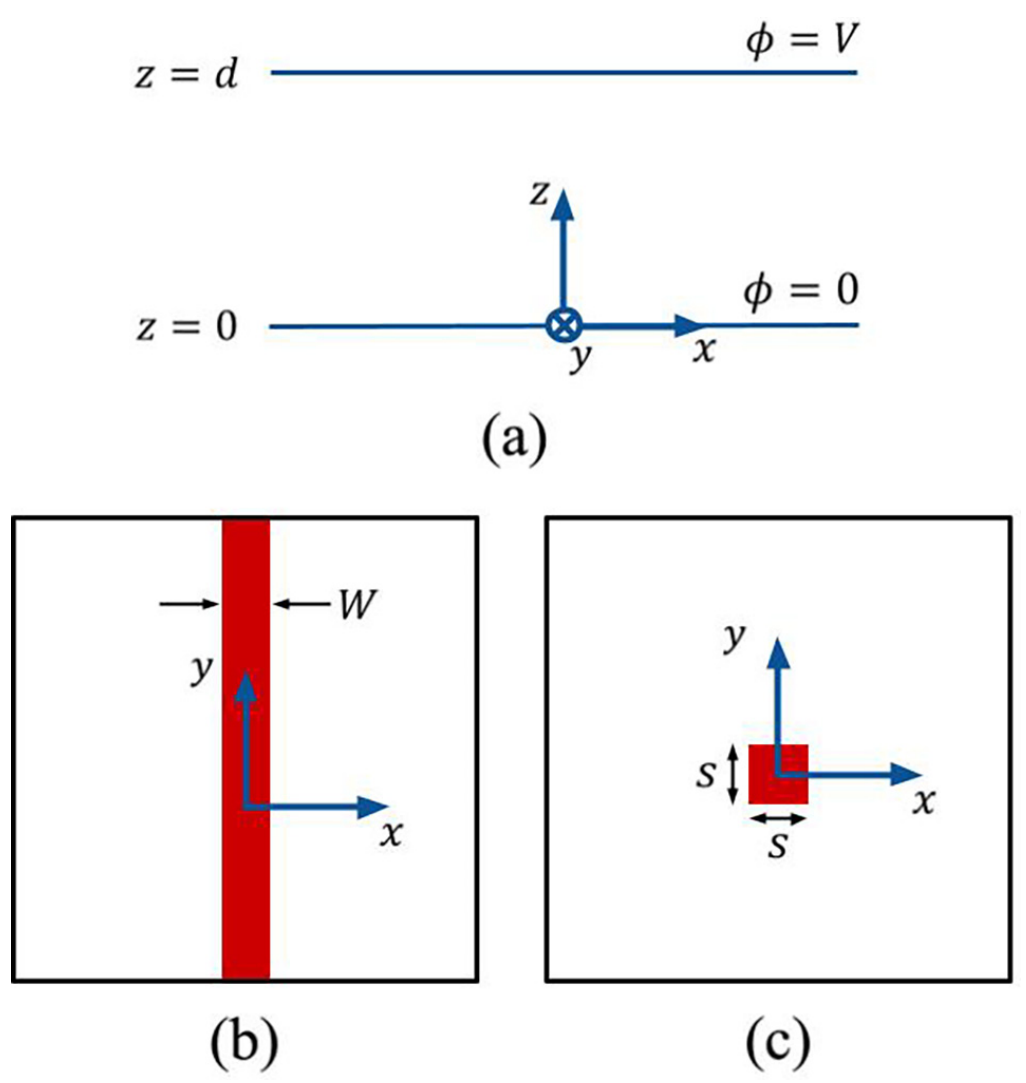

**FIG. 1.** (a) A planar diode of gap spacing $d$ and gap voltage $V$. (b) Emitting stripe of width $W$ on the cathode ($z = 0$). (c) Square emitting tile of size $s$ on the cathode. This paper considers the limits $W \to 0$ and $s \to 0$, corresponding to an electron sheet and a line charge, respectively.

Umstattd and Luginsland[10] considered a similar 2D problem but allowed the entire emitting strip to satisfy the space-charge-limited condition, i.e., the electric field on the cathode surface equals zero everywhere on the emitting stripe [Fig. 1(b)]. Their simulation study revealed several important features.

(a) The emitted current density profile exhibits a wing-like structure at the edges of the emitting stripe where the local current density is significantly higher than the 1D CL value, Eq. (1.1), due to the lack of space charge in the region adjoining the emitting stripe.
(b) The significant increase in the edge current may compensate for the non-emitting regions to the extent that if only 20% of the cathode surface is actively emitting (with the remaining 80% of the cathode surface non-emitting) the cathode may still deliver 80% of the 1D CL current, as if the entire cathode were emitting.
(c) The edge effect in (b) is most pronounced for emitting stripes with narrower width.
(d) The emitted current density's wing-like structure is independent of the longitudinal applied magnetic field, similar to the conclusions of Luginsland *et al.*[8] Thus, in an analytic theory, for simplicity, an infinite longitudinal magnetic field may be assumed to restrict electron motion to one direction, and this paper will adopt this simplifying assumption.

Chernin *et al.*[13] and Jassem *et al.*[14] used both a semi-analytical method and the MICHELLE particle-in-cell code[30] to study non-uniform emission from a thermionic cathode that underwent a transition from the temperature-limited regime to the space-charge-limited regime as the cathode surface temperature was raised. They considered emission patches in the form of stripes [1D, Fig. 1(b)] and of square tiles [2D, Fig. 1(c)], respectively. They modeled realistic levels of emission non-uniformity through 1D[13] and 2D[14] variations of the work functions on the cathode surface, where the work function distributions were obtained from electron backscatter diffraction measurements on a tungsten dispenser cathode.[31] They found excellent agreement between the semi-analytical formulation and the MICHELLE code results for all electron flow regimes: temperature-limited, space-charge-limited, and the transition between them, even when the aspect ratio such as $d/W$ is of order $10^3$, which was far beyond the range of Eq. (1.2) and of the simulations by Umstattd *et al.*[10] and Luginsland *et al.*[8] Their study revealed the following additional features.[13,14]

(e) The emitting patches do not emit independently. The anode current predominantly originates from the regions of the lowest work function, even though such regions constitute of only 18% of the total cathode area, as in the experimental data that they used.
(f) For a fixed work function distribution, as the emitting patch size shrinks, the contributions from the lowest work function regions become even more dominant at high temperatures, because the edge effect is roughly measured by the ratio of the circumference to the area of an emission patch, and this ratio increases as the patch size decreases.

It is, therefore, of substantial interest to establish scaling laws both in 2D and 3D for emission patches with very small emission size. This paper considers this problem, since the analytic scaling laws, (1.2) and (1.3), are not valid when the emission patch size, $W$ or $R$, is very small compared with the gap separation $d$. In addition, edge emission from small patches is an important contributor to the anode current [cf. point (f)]. Thus, for 2D, we consider an isolated emission stripe in a planar cathode with a vanishingly small stripe width $W$ [Fig. 1(b)], so that the emission current forms an electron sheet whose density profile is a delta function in $x$. For 3D, we consider an isolated emission square tile, as shown in Fig. 1(c), (or emission circular patch) with a vanishingly small tile size $s$ (or patch radius $R$) so that the emission current forms a line charge whose density profile is a delta function both in $x$ and in $y$. Since lateral motion of the electrons has been shown to be unimportant,[13,14] we shall assume an infinite longitudinal magnetic field so that electron motion is confined to the $z$-direction [see also point (d) above].

The assumption of an emission current density profile in the form of a delta function, in both a 2D and 3D geometry [Figs. 1(b) and 1(c)], conveniently bypasses the vexing question concerning whether, and where, the electron emission is space-charge-limited in an extremely small emission patch. It also bypasses the geometrical question of whether this very small emission patch is a square tile or circular in shape. By solving the Poisson equation for delta-function emission current density profiles, our results are independent of the cathode surface properties. We shall, however, compare this new theory with some key results of our previous work on non-uniform emission, as summarized above.

For the sheet and line current problems, we have formulated the limiting current in terms of an integral equation. We have solved this equation iteratively, in the manner given in Sec. II. The numerical results are presented in Sec. III. To validate this approach, since limiting current in a diode has not been previously formulated in terms of an integral equation to our knowledge, we first present in Subsec. III A the numerical results for the classical 1D CLL using this integral equation approach. We also include Jaffe's extension to a constant, nonzero initial velocity of the emitted electrons.[32] In Subsec. III B, the results for the 2D limiting current of an electron sheet are presented. We show that a solution exists if and only if we assume the sheet electrons are emitted with a non-zero initial velocity. The solution turns out to be in qualitative agreement with Chernin's 2D study of thermionic cathodes[13] (as will be shown in Sec. IV). Subsection III C considers the limiting current for an electron line charge. In this case, we show that there is no non-trivial solution regardless of the emission velocity of the electrons. An interpretation of this null result is given. In Sec. IV, we will further show that this null result may actually be anticipated from the data of Jassem's 3D study of thermionic cathodes;[14] we also show how point (f) above is reconciled with this null result. Section V concludes this study. The main results are given in the main text; the detailed derivations, the mathematical proof of nonexistence of solution, and the discussions of some mathematical subtleties, are given in the Appendixes.

## II. FORMULATION

We consider a planar diode with gap separation $d$ and gap voltage $V$ [Fig. 1(a)]. An infinite magnetic field in the z-direction is assumed so that all electron motions are restricted to the z-direction. The charge density, $-\rho(x,y,z)$, and the electrostatic potential $\phi(x,y,z)$ satisfy the Poisson equation ($\rho \geq 0$),

$$\nabla^2\phi(x,y,z) = \frac{\rho(x,y,z)}{\epsilon_0} \tag{2.1}$$

with the boundary conditions

$$\phi(x,y,0) = 0, \quad \phi(x,y,d) = V. \tag{2.2}$$

We write $\phi(x,y,z)$ as

$$\phi(x,y,z) = Vz/d + \psi(x,y,z), \tag{2.3}$$

which is a superposition of the vacuum potential, $Vz/d$, and the space charge potential, $\psi(x,y,z)$, that satisfies the Poisson equation,

$$\nabla^2\psi(x,y,z) = \frac{\rho(x,y,z)}{\epsilon_0} \tag{2.4}$$

with the grounded boundary conditions at $z = 0$ and $z = d$,

$$\psi(x,y,0) = 0, \quad \psi(x,y,d) = 0. \tag{2.5}$$

For the classical 1D Child–Langmuir problem, the magnitude of the charge density, $\rho(x,y,z)$, depends only on $z$, and the space charge may be considered as a superposition of electron sheets within the gap [Fig. 2(a)]. The potential $\psi(x,y,z)$ due to a typical electron sheet, located at $z = z_c$ [Fig. 2(a)], may readily be obtained; it is simply the Green's function, $G(z, z_c)$, for Eqs. (2.4) and (2.5), derived in Sec. III. It is important to note that $G(0, z_c) = G(d, z_c) = 0$, since the homogeneous boundary condition (2.5) is satisfied. Superposition of this Green's function yields the space charge field due to $\rho(z) \equiv \rho(0,0,z)$ for the classical 1D Child–Langmuir problem. We further assume that all electrons are emitted from the cathode with the same velocity in the z-direction with energy, $E_{in}$. The electron velocity $v(z)$ at a position $z$ is given by $mv^2/2 = E_{in} + e\phi(z)$, yielding

$$\rho(z) = \frac{J}{v(z)} = J/\left[(2/m)(E_{in} + e\phi(z))\right]^{1/2}, \tag{2.6}$$

where $\phi(z) \equiv \phi(0,0,z)$ and $J\,(> 0)$ is the current density along the z-axis, (x,y) = (0,0). Note that $J$ is a constant, independent of $z$. Evaluating Eq. (2.3) at (x,y) = (0,0), and noting the remarks following Eq. (2.5), we arrive at the integral equation for $\phi(z)$ in an alternate derivation of the classical 1D Child–Langmuir law,

$$\phi(z) = Vz/d + \int_0^d dz_c \frac{J}{\left[(2/m)\left(E_{in} + e\phi(z_c)\right)\right]^{1/2}} G(z, z_c). \tag{2.7}$$

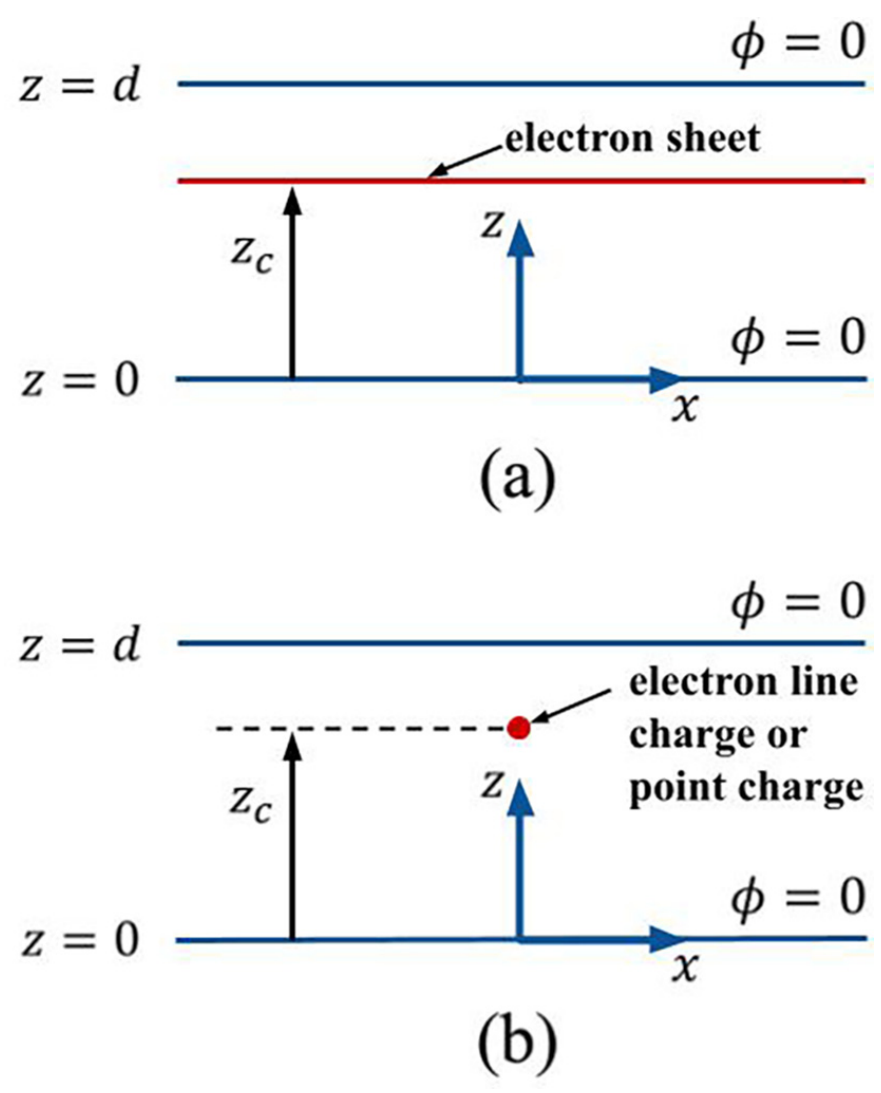

**FIG. 2.** (a) An electron sheet and (b) a line charge or a point charge between two grounded plates. The Green's function of the space charge potential at (x,y,z) = (0,0,z) is constructed to calculate the limiting current in 1D using (a) in Subsec. III A, and in 2D and 3D using (b) in Subsecs. III B and III C, respectively.

The CLL limiting current density, in this formulation, is the value of $J$ beyond which there is no solution to the integral equation (2.7), under the assumption $E_{in} = 0$. Jaffe extends the CLL to $E_{in} > 0$.[32]

For the 2D problem, we assume that the emission region is a stripe of a vanishingly small width, $W$ [Fig. 1(b)]. We may similarly construct the integral equation for $\phi(z) \equiv \phi(0,0,z)$. This electron sheet may be considered as a superposition of line charges located at (x,y) = (0,0). The Green's function, $G(z, z_c)$ for Eqs. (2.4) and (2.5), due to a representative line charge located at $z = z_c$ of unit line charge density [Fig. 2(b)], may be obtained from the image charge method. Note that Eqs. (2.1)–(2.7) still apply for this electron sheet problem. The last statement, likewise, also applies for the 3D problem where $\rho(x, y, z)$ in Eq. (2.1) represents a line charge located at $(x,y) = (0,0)$, as shown in Fig. 1(c) in which the tile size $s \to 0$.

Defining the dimensionless variables $\overline{\phi} = \phi/V$, $\overline{z} = z/d$, $\overline{z}_c = z_c/d$, the general integral equation (2.7) becomes

$$\overline{\phi}(\overline{z}) = \overline{z} + K\int_0^1 \frac{d\overline{z}_c}{\left(\overline{\phi}(\overline{z}_c)+\Delta\right)^{1/2}}\overline{G}(\overline{z},\overline{z}_c), \quad 0 \le \overline{z} \le 1, \tag{2.8}$$

$$\Delta = E_{in}/eV, \tag{2.9}$$

where $\Delta$ is the dimensionless parameter measuring the injection energy of the mono-energetic electrons, and $K(\ge 0)$ is the dimensionless parameter proportional to the emission current (which is equal to the anode current for the present model of a mono-energetic electron stream). Note that the integral in Eq. (2.8) is always negative, as it represents the potential depression due to some unit electron charge inside two grounded plates (Fig. 2). The limiting current is given by the maximum value of $K$ beyond which there is no solution to Eq. (2.8). If $K = 0$, Eq. (2.8) yields the vacuum field solution, $\overline{\phi}(\overline{z}) = \overline{z}$, as expected. For a small value of $K$, we expect that Eq. (2.8) may be solved iteratively, starting with this vacuum solution. The approximate solution after the $k$-th iteration is then given by

$$\overline{\phi}^{(k)}(\overline{z}) = \overline{z} + K\int_0^1 \frac{d\overline{z}_c}{\left(\overline{\phi}^{(k-1)}(\overline{z}_c)+\Delta\right)^{1/2}}\overline{G}(\overline{z},\overline{z}_c),$$

$$k = 1, 2, 3, \ldots, \quad \overline{\phi}^{(0)}(\overline{z}) = \overline{z}. \tag{2.10}$$

At a specified value of $\Delta$, we consider that the limiting current (maximum value of $K$) is reached if after some $k$-step iterations, $\overline{\phi}^{(k)}(\overline{z})$ first becomes complex at any value of $\overline{z}$ between (0,1). Since Eq. (2.8) is real, this condition is equivalent to $\overline{\phi}(\overline{z}) + \Delta < 0$ after some $k$-iterations at any value of $\overline{z}$ between (0,1). Note further that we implicitly equate *non-convergence* of the iterative algorithm of Eq. (2.10) with *nonexistence* of a solution, but we have not proven it. However, the fact that we are able to recover the classical CLL, including Jaffe's extension,[32] give us some confidence in its validity. In Sec. III, we present the limiting currents, thus obtained from this iterative method in Subsecs. III A, III B, and III C, respectively, for the three cases: the classical 1D CLL, electron emission in the form of a thin sheet, and electron emission in the form of a thin line.

## III. THE ITERATIVE SOLUTIONS

This section outlines the iterative solutions for the three cases listed in the last sentence of Sec. II. The details are given in the Appendixes.

### A. The classical 1D Child–Langmuir law

In this case, $\rho(x, y, z) = \rho(z)$ in Eqs. (2.1) and (2.4). The Green's function $G(z, z_c)$ for Eqs. (2.4) and (2.5) is the electrostatic potential due to a representative electron sheet charge, located at $z = z_c$ of unit surface charge density [Fig. 2(a)]. It satisfies

$$\frac{d^2G(z,z_c)}{dz^2} = \frac{1}{\epsilon_0}\delta(z - z_c), \tag{3.1}$$

where $\delta$ denotes the Dirac delta function and $G(0, z_c) = G(d, z_c) = 0$. We have denoted $\psi(0, 0, z) = G(z, z_c)$ here and henceforth in Sec. III. The solution to Eq. (3.1) is readily shown to be

$$G(z,z_c) = \begin{cases} -\dfrac{(d-z_c)z}{\epsilon_0 d}, & z \le z_c, \\ \dfrac{z_c(z-d)}{\epsilon_0 d}, & z > z_c, \end{cases} \tag{3.2}$$

which is plotted in Fig. 3. Equation (2.7) then reads

$$\begin{aligned}\phi(z) &= Vz/d + \int_0^d [\rho(z_c)\,dz_c]\,G(z,z_c) \\ &= Vz/d + \int_0^z dz_c \frac{J}{[(2/m)(E_{in}+e\phi(z_c))]^{1/2}}\left[\frac{z_c(z-d)}{\epsilon_0 d}\right] \\ &\quad + \int_z^d dz_c \frac{J}{[(2/m)(E_{in}+e\phi(z_c))]^{1/2}}\left[-\frac{(d-z_c)z}{\epsilon_0 d}\right], \\ &\quad 0 \le z \le d. \end{aligned} \tag{3.3}$$

Its normalized form, Eq. (2.8), becomes

$$\overline{\phi}(\overline{z}) = \overline{z} + K_1\left\{\int_0^{\overline{z}} \frac{d\overline{z}_c}{\left(\overline{\phi}(\overline{z}_c)+\Delta\right)^{1/2}}[\overline{z}_c(\overline{z}-1)] + \int_{\overline{z}}^1 \frac{d\overline{z}_c}{\left(\overline{\phi}(\overline{z}_c)+\Delta\right)^{1/2}}[-\overline{z}(1-\overline{z}_c)]\right\}, \quad 0 \le \overline{z} \le 1, \tag{3.4}$$

where

$$K_1 = \frac{(4/9)J}{J_{CL}} > 0, \tag{3.5}$$

and $J_{CL}$ is the 1D classical Child–Langmuir current density, Eq. (1.1).

We use the iterative scheme, Eq. (2.10), on Eq. (3.4). We consider the limiting current (maximum value of $K_1$) is reached if after some $k$-step iterations, $\overline{\phi}^{(k)}(\overline{z})$ first becomes complex at any value of $\overline{z}$ between (0,1). For $\Delta = 0$, we have found agreement within 0.1% between our numerical results for the maximum value of $J$ with the classical 1D CLL, Eq. (1.1). For nonzero $\Delta$, the numerical scheme (2.10) yields the maximum value $J = J(1)$, which is shown in Fig. 4 for some test cases at very low and high values of $\Delta$. Also shown in Fig. 4 is Jaffe's analytic formula for nonzero $\Delta$,[32]

$$\frac{J(1)}{J_{CL}} = \left[(1+\Delta)^{1/2} + \Delta^{1/2}\right]^3. \quad \text{(Jaffe)} \tag{3.6}$$

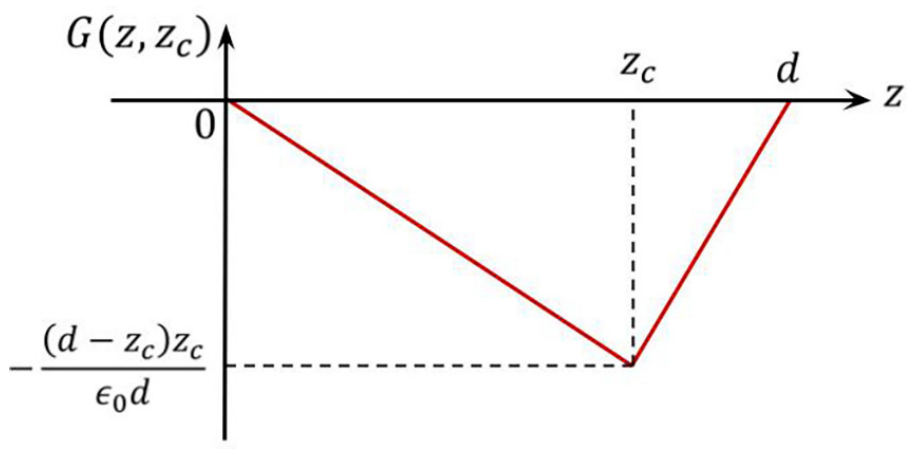

FIG. 3. The Green's function $G(z, z_c)$ for the 1D geometry (the classical Child–Langmuir law).

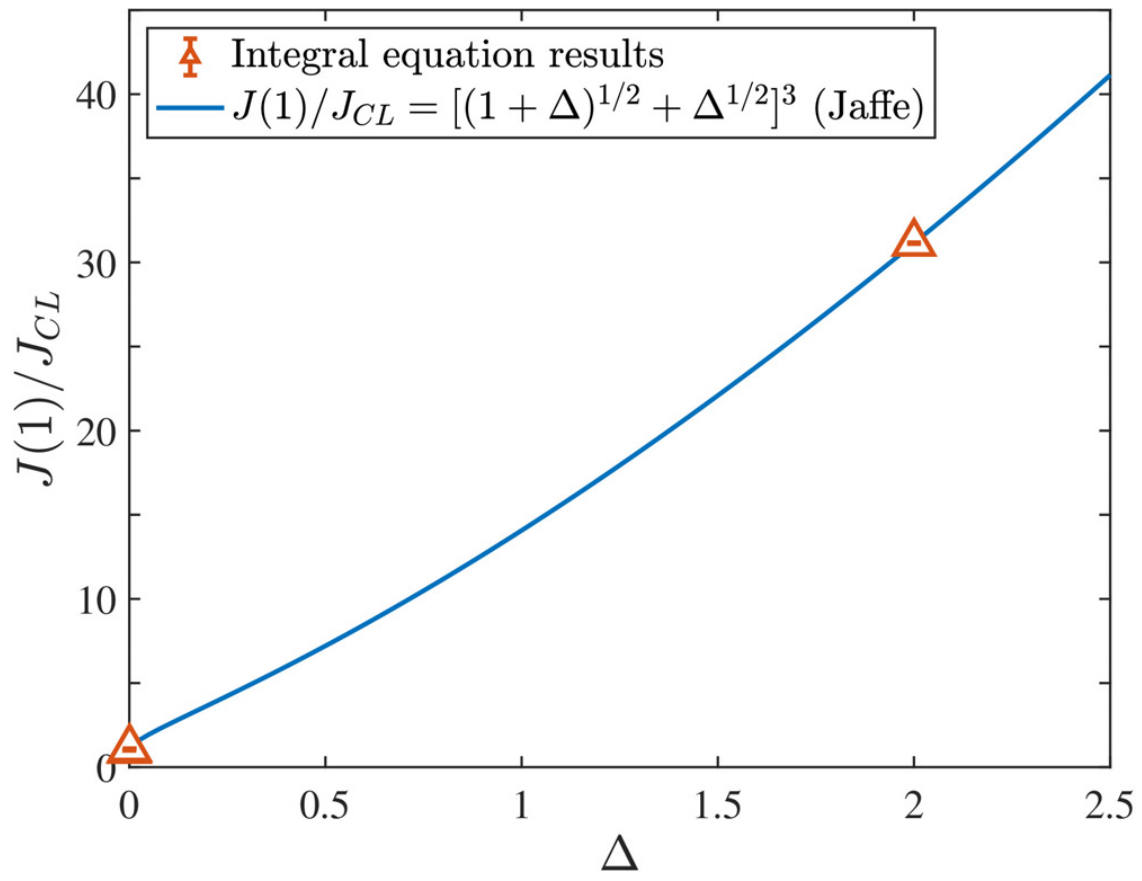

**FIG. 4.** The 1D Child–Langmuir Law from solutions of the integral equation (3.4) for $\Delta = 0$, $10^{-3}$, 2 (triangles, with numerical error less than 0.5%). The solid line shows Jaffe's formula for nonzero initial emission energy, $\Delta = E_{in}/\mathrm{eV} \geq 0$. The numerical values of $J(1)/J_{CL}$ at $\Delta = 0$, $10^{-3}$, 2 are, respectively, 1, 1.0995, 31.1448, according to Jaffe.

Figure 5 shows the potential profiles $\overline{\phi}(\overline{z})$ obtained from the iterative scheme, for $\Delta = 0$, and $\Delta = 2$, at the maximum value of $K_1$. These curves are indistinguishable from the analytic solution. Note that in Fig. 5, $\overline{\phi}(\overline{z})$ does not reach the value $-\Delta$ at its minimum at the limiting current, a well-known result for nonzero $\Delta$,[32] even though we use $\overline{\phi}(\overline{z}) + \Delta < 0$ after some k-iterations at any value of $\overline{z}$ between (0,1) as the condition for nonexistence of solution. Appendix A discusses the convergence of the iterative solution to Eq. (3.4) and related numerical issues. The numerical results shown in Figs. 4 and 5 gave us confidence in using the same iterative scheme, (2.10), on the electron sheet and electron line charge problems, at least for low values of $K$.

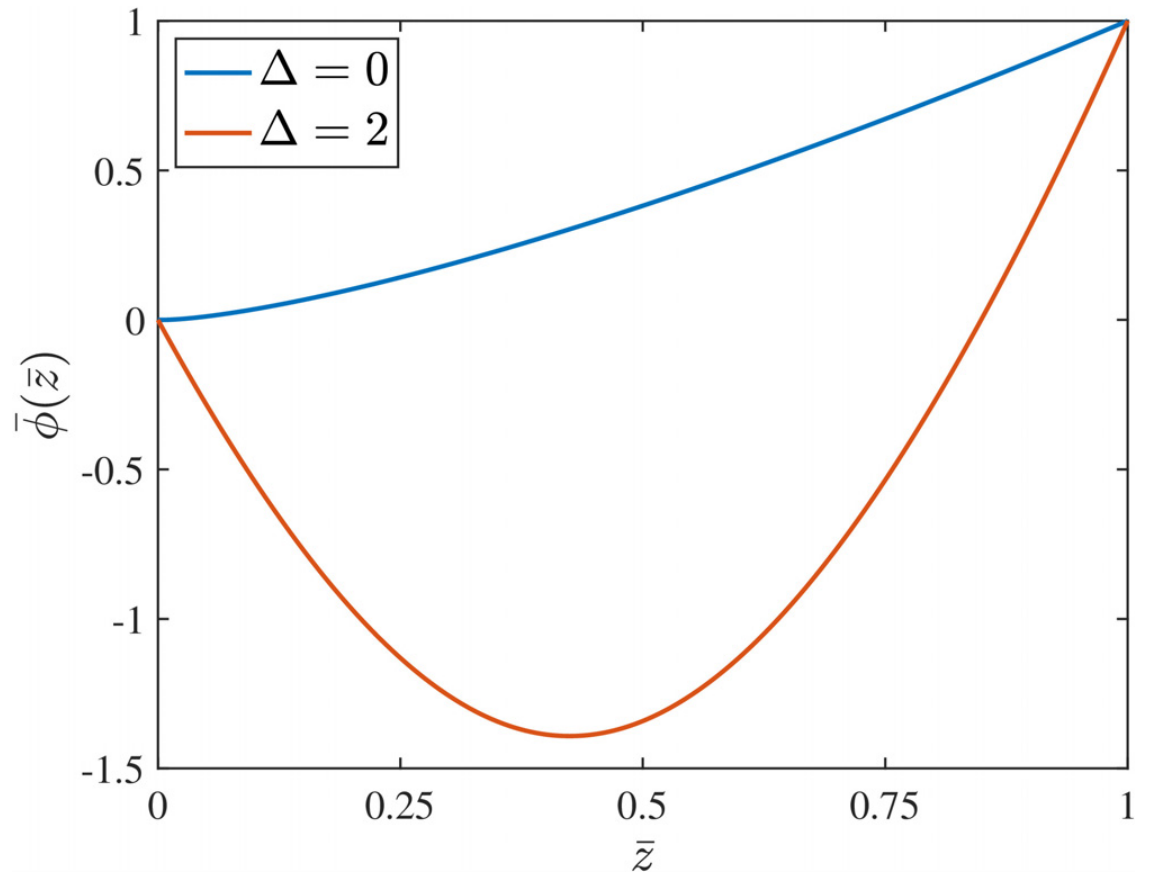

**FIG. 5.** Potential profiles at the limiting current in 1D for $\Delta = 0$ and $\Delta = 2$, from the numerical solution to the integral equation (3.4). These two curves are indistinguishable from the analytic theory.

## B. Maximum sheet current

In this case, $\rho(x,y,z) = \sigma(z)\delta(x)$ in Eqs. (2.1) and (2.4), where $\sigma(z) > 0$ is the magnitude of the surface charge density, in $\mathrm{C/m^2}$, for a current sheet of vanishingly small thickness [Fig. 1(b)]. The Green's function to Eqs. (2.4) and (2.5) is the electrostatic potential due to a line charge of unit line charge density located at $(x,z) = (0,z_c)$, $z_c \in (0,d)$ that satisfies [cf. Fig. 2(b)]

$$\nabla^2\psi(x,y,z) = \frac{1}{\epsilon_0}\delta(z-z_c)\delta(x) \tag{3.7}$$

and the homogeneous boundary condition, Eq. (2.5). The solution $\psi(x,y,z)$ within the gap [Fig. 2(b)] may be readily obtained by summing the electrostatic potential due to the infinite series of image line charges located at $z = (2nd + z_c)$, $n = \pm 1, \pm 2, \ldots$ From $\psi(0,0,z)$, which we denote as $G(z,z_c)$ as in Eq. (3.1), we obtain

$$\begin{aligned} G(z,z_c) &= \frac{1}{2\pi\epsilon_0}\sum_{n=-\infty}^{\infty}\ln\left[\frac{|z-(2nd+z_c)|}{|z+(2nd+z_c)|}\right] \\ &= \frac{1}{2\pi\epsilon_0}\ln\left[\frac{\left|\sin\left(\frac{\pi}{2}(-\overline{z}+\overline{z}_c)\right)\right|}{\sin\left(\frac{\pi}{2}(\overline{z}+\overline{z}_c)\right)}\right], \\ &0 \leq z \leq d,\ 0 < z_c < d. \end{aligned} \tag{3.8}$$

Note that $G(0,z_c) = G(d,z_c) = 0$, since the homogeneous boundary condition (2.5) is satisfied. The last equality in Eq. (3.8) may be established using the identity,

$$\prod_{m=-\infty}^{\infty}\frac{m+a}{m+b} = \frac{\sin(\pi a)}{\sin(\pi b)}. \tag{3.9}$$

The normalized integral equation (2.8) then reads

$$\overline{\phi}(\overline{z}) = \overline{z} + K_2\int_0^1 \frac{d\overline{z}_c}{\left(\overline{\phi}(\overline{z}_c)+\Delta\right)^{1/2}}\ln\left[\frac{\left|\sin\left(\frac{\pi}{2}(-\overline{z}+\overline{z}_c)\right)\right|}{\sin\left(\frac{\pi}{2}(\overline{z}+\overline{z}_c)\right)}\right],$$

$$0 \leq \overline{z} \leq 1, \tag{3.10}$$

$$K_2 = (2/9\pi)(M_2/d)/J_{CL}, \tag{3.11}$$

where $M_2$ (>0, in A/m) is the current carried by the electron sheet per unit length in $y$ [Fig. 1(b)]. Note that $M_2 = \sigma(z)v(z)$ is a positive constant.

As in Subsec. III A, we numerically solve Eq. (3.10) using the iterative scheme (2.10). For the case of zero injection energy ($\Delta = 0$), we only find the null solution for $K_2$, i.e., there is no non-zero value of $K_2$, no matter how small, for which Eq. (3.10) has a solution for $\overline{\phi}(\overline{z})$ when $\Delta = 0$. We obtain this surprising result after considerable numerical effort, as outlined in Appendix B. The mathematical proof is given in Appendix C. Appendix B describes two numerical methods that we have used to validate each other.

For nonzero $\Delta$, the iterative scheme converges if $K_2$ is below a critical value, which is plotted in Fig. 6. This figure shows that this critical $K_2$ is numerically quite small. We argue in Sec. IV that these values of $K_2$ are consistent with previous studies of 2D emission stripes on thermionic cathodes that used realistic work function distributions. Note that by finding non-zero solutions for delta-function current density profiles, we have demonstrated that it is possible in principle to exceed $J_{CL}$ locally by an arbitrary large factor for nonzero emission velocity.

### C. Maximum line current

In this case, $\rho(x, y, z) = \lambda(z)\delta(x)\delta(y)$ in Eqs. (2.1) and (2.4), where $\lambda(z) > 0$ is the magnitude of the line charge density, in C/m, for an electron line charge of vanishingly small cross section [Fig. 1(c)]. The Green's function to Eqs. (2.4) and (2.5) is the electrostatic potential due to a unit point charge located at $(x, y, z) = (0, 0, z_c)$, $z_c \in (0, d)$ that satisfies [Fig. 2(b)]

$$\nabla^2\psi(x, y, z) = \frac{1}{\epsilon_0}\delta(z - z_c)\delta(x)\delta(y) \tag{3.12}$$

and the homogeneous boundary condition, Eq. (2.5). The solution $\psi(x, y, z)$ within the gap [Fig. 2(b)] may also be obtained by summing the electrostatic potential due to the infinite series of image point charges located at $z = (2nd + z_c)$, $n = \pm1, \pm2, \ldots$. From $\psi(0, 0, z)$, in which we obtain

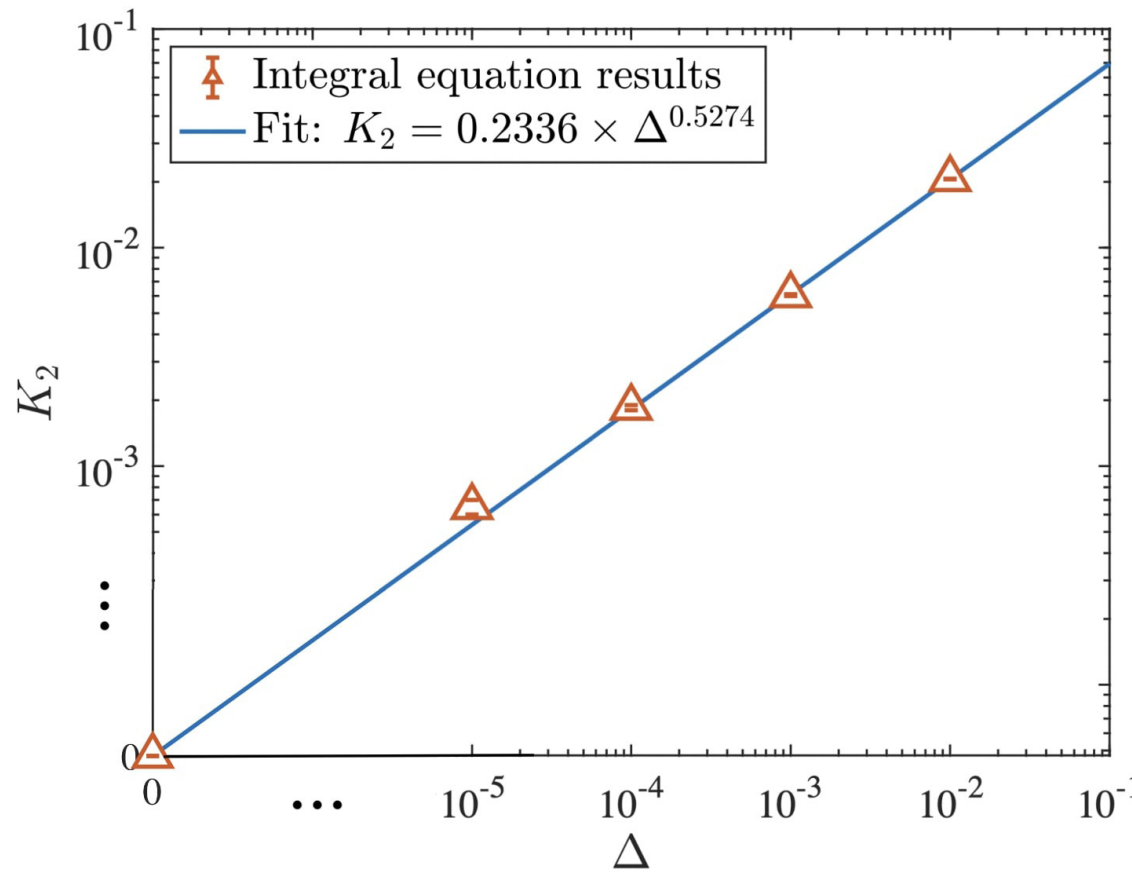

**FIG. 6.** The normalized 2D limiting current ($K_2$) on an electron sheet as a function of $\Delta$ according to the integral equation formulation (triangles). The solid line of best fit (having both R-square and adjusted R-square values of 0.9999) is added for visual convenience. The numerical values of $K_2$ at $\Delta = 0$, $10^{-5}$, $10^{-4}$, $10^{-3}$, and $10^{-2}$ are, respectively, 0, $6.5 \times 10^{-4}$, $1.85 \times 10^{-3}$, $6.05 \times 10^{-3}$, and $2.06 \times 10^{-2}$, whose respective errors are 0, $5 \times 10^{-5}$, $5 \times 10^{-5}$, $5 \times 10^{-5}$, and $5.315 \times 10^{-5}$.

$$G(z, z_c) = \frac{1}{4\pi\epsilon_0}\Bigg\{-\frac{1}{|z - z_c|} + \frac{1}{|z + z_c|} + \sum_{n=1}^{\infty}\left[\frac{1}{|z + (2nd + z_c)|} - \frac{1}{|z - (2nd + z_c)|}\right] - \sum_{n=1}^{\infty}\left[\frac{1}{|z + (2nd - z_c)|} - \frac{1}{|z - (2nd - z_c)|}\right]\Bigg\}.$$
$$0 \le z \le d, \quad 0 < z_c < d. \tag{3.13}$$

The normalized integral equation (2.8) then reads

$$\overline{\phi}(\overline{z}) = \overline{z} + K_3\int_0^1 \frac{d\overline{z}_c}{\left(\overline{\phi}(\overline{z}_c) + \Delta\right)^{1/2}}\overline{H}(\overline{z}, \overline{z}_c), \quad 0 \le \overline{z} \le 1, \tag{3.14}$$

$$\overline{H}(\overline{z}, \overline{z}_c) = -\frac{1}{|\overline{z} - \overline{z}_c|} + \frac{1}{|\overline{z} + \overline{z}_c|} + \sum_{n=1}^{\infty}\left[\frac{1}{|\overline{z} + (2n + \overline{z}_c)|} - \frac{1}{|\overline{z} - (2n + \overline{z}_c)|}\right] - \sum_{n=1}^{\infty}\left[\frac{1}{|\overline{z} + (2n - \overline{z}_c)|} - \frac{1}{|\overline{z} - (2n - \overline{z}_c)|}\right], \tag{3.15}$$

$$K_3 = \left(I_3/d^2\right)/(9\pi J_{CL}). \tag{3.16}$$

In Eq. (3.16), $I_3$ (>0, in A) is the current carried by the electron line charge [Fig. 1(c)]. Note that $I_3 = \lambda(z)v(z)$ is a positive constant.

It is easy to see that there is no solution to Eq. (3.14) for $K_3 \neq 0$ because the singularity at $\overline{z}_c = \overline{z}$ in $\overline{H}(\overline{z}, \overline{z}_c)$ is so strong that the integral (3.14) always diverges for any $\overline{z}$, $0 < \overline{z} < 1$. It then immediately follows that only the null solution for the total line charge current exists, regardless of the electrons' emission velocity. This result might have been anticipated from the mathematical idealization of a line charge, on which the electrostatic potential approaches negative infinity, logarithmically. This infinitely large negative potential barrier prevents electron travel toward the anode regardless of the electron's initial velocity. We conjecture that it is a similar negative potential on the cathode surface that causes the null solution for the electron sheet problem in Subsec. III B when the injection energy is zero. In fact, Appendix C shows that the nonexistence of a solution for the electron sheet geometry (with $\Delta = 0$) first appears at $z = 0$. On the other hand, in Subsec. III A, such a potential barrier is absent in the 1D classical Child–Langmuir problem even with $\Delta = 0$. Mathematically, the Green's function $G(z, z_c)$ is finite at $z_c = z$ in Subsec. III A, but is negative infinite in Subsecs. III B and III C.

Despite the zero current limit for a line current, we note that in a realistic thermionic cathode with patchy emission on the cathode surface, it is the tiny emitting patches that carry the greatest fraction of anode current, as pointed out in item (f) in Sec. I. In Sec. IV, we attempt to resolve this paradox by a comparison with previous analytic theory and simulation, which use realistic work function distributions to account for nonuniform electron emission on a cathode.

## IV. COMPARISON WITH PREVIOUS THEORIES ON PATCHY EMISSION

This section shows that the seemingly surprising results of Subsecs. III B and III C are indeed consistent with Umstattd and Luginsland,[10] Chernin *et al.*,[13] and Jassem *et al.*[14] Consider first electron emission from a single *isolated* patch of area $A_e$. Let $I_A$ be the total current reaching the anode from this isolated patch, and $J_A \equiv I_A/A_e$ be the anode current density resulting from this isolated emitting patch. We may write, in general,

$$J_A = J_{CL} + \Delta J, \tag{4.1}$$

since it was established that the anode current density from a finite patch may exceed the 1D Child–Langmuir value by $\Delta J$, as shown in Eqs. (1.2) and (1.3) for instance. Suppose that there are $N$ such emitting patches on this cathode and that the remaining area on this cathode, designated as $A_n$, is non-emitting. Then, the total current reaching the anode is, assuming that each emitting patch remains isolated,

$$I(2) = NA_e(J_{CL} + \Delta J). \tag{4.2}$$

The total anode current according to the 1D CLL is, assuming that the entire cathode is emitting,

$$I(1) = (NA_e + A_n)J_{CL}. \tag{4.3}$$

For $I(2)$ to approach $I(1)$, the following condition needs to be satisfied, upon comparing Eqs. (4.2) and (4.3),

$$\frac{\Delta J}{J_{CL}} > \frac{A_n}{NA_e}, \tag{4.4}$$

since each emitting patch may no longer be isolated from its surroundings on a real cathode [cf. point (e) in Sec. I].

The inequality (4.4) has an interesting interpretation. Its right-hand side (RHS) is simply the ratio of the non-emitting area to the highly emitting area on the entire cathode. This ratio could be large, ~4 for example, if the non-emitting area is four times the actively emitting area; i.e., only 20% of the cathode surface is actively emitting and the remaining 80% is non-emitting. However, the anode current may still approach Eq. (4.3) if each emitting patch produces a sufficiently large $\Delta J$ to compensate for the non-emitting regions. An emitting patch whose size is much smaller than the AK spacing could satisfy Eq. (4.4). This was indeed a major discovery by Umstattd and Luginsland,[10] reenforced by Chernin *et al.*[13] and Jassem *et al.*[14] These authors' numerical calculations all show that a cathode with only 20%,[10] or even 5% (Ref. 13) of tiny, actively emitting patches may deliver up to 80% of the 1D CL current for the entire cathode.

We now compare the key unexpected results of the present paper with some specific examples in previous studies of patchy emissions on a thermionic cathode. In this comparison, the AK gap spacing was fixed at $d = 381\ \mu$m, and the gap voltage was $V = 179.5$ V. For these parameters, $J_{CL} = 4.2\ \mathrm{A/cm^2}$ (including the small correction due to the finite cathode temperature[3,33]). Chernin *et al.*[13] considered a two-stripe model in a 2D theory, one stripe is emitting with a work function of 2.1 eV, and the neighboring stripe is non-emitting. The total width (in $y$, in Chernin's notation here) of both stripes is fixed at $p = 20\ \mu$m, and the work function distribution is periodic in $y$ with period $p$. Within each period, the width ($W$) of the emitting stripe ranges from 5% to 100% of the full period $p$. At a cathode temperature $T = 1400$°C, both Chernin's semi-analytic theory and the MICHELLE code yield $I(2)$ to be within 90% of $I(1)$ for all $W > 0.1p = 2\ \mu$m. (See Fig. 10 of Chernin.[13]) Thus, a 2 $\mu$m width of an emitting stripe is very small compared with $d = 381\ \mu$m, to a large extent such an emission stripe might be modeled as a delta-function distribution, as done in Subsec. III B. To within 10% accuracy, for $W > 2\ \mu$m, we then obtain, for this example,

$$\frac{J(2)}{J_{CL}} \approx \frac{p}{W}, \quad 0.1p < W < p. \tag{4.5}$$

For an emission stripe of width $W$, the sheet current density is $M_2 = W \times J(2) = pJ_{CL}$, yielding the normalized sheet current parameter [cf. Eq. (3.11)]

$$K_2 = \frac{2}{9\pi} \times \frac{M_2}{dJ_{CL}} = \frac{2}{9\pi}\frac{p}{d} = 0.00371. \tag{4.6}$$

Note that a cathode temperature of 1400 °C (~0.14 eV) roughly corresponds to the initial emission velocity parameter $\Delta = 0.14\ \mathrm{V}/179.5\ \mathrm{V} = 0.000\,78$, which is between the range of $\Delta = 0.0001$ and $\Delta = 0.001$ (Fig. 6). Note further that Eq. (4.6) gives a value of $K_2$ between 0.001 85 and 0.006 05, the latter two numbers being the critical values of $K_2$ corresponding to $\Delta = 0.0001$ and $\Delta = 0.001$, respectively (cf. Fig. 6). *Thus, our unexpected, new result on current sheet as given in Subsec. III B is in fact consistent with previous studies of 2D emitting stripes on thermionic cathodes that used realistic work function distributions.*

Let us now turn to the 3D extension of non-uniform emission on a thermionic cathode with the same $d = 381\ \mu$m and $V = 179.5$ V as in the preceding paragraph, also used by Jassem *et al.*[14] In one example that is most relevant to the study of a line charge given in Subsec. III C, Jassem considers a work function distribution that is periodic both in $x$ and in $y$ on the cathode surface, with equal period $p$. The simulated cathode surface, of area $p \times p$, is subdivided into 256 square tiles, each tile having an edge of length $s$ [Fig. 1(c)] so that $p = 16s$. The results for $s = 0.3125,\ 2.5,\ 5,$ and $10\ \mu$m are shown in Fig. 5 of Jassem.[14] At a sufficiently high cathode temperature, such as 1200 °C, this figure shows that the anode current for all values of $s$ is within three percent of the 1D Child–Langmuir value, as if the entire cathode were emitting. Note that this anode current predominantly comes from the 46 tiles of the lowest work function [1.61 eV, see Fig. 4(e) of Jassem[14]], which constitute of only 46/256 = 17.97% of the cathode area (see Table I of Jassem). Thus, analogous to Eq. (4.5), these numerical results suggest, for this example,

$$\frac{J(3)}{J_{CL}} \approx \frac{p^2}{46 \times s^2} = 5.57. \tag{4.7}$$

Equation (4.7) implies that the total anode current due to each square tile of size $s$ is approximately $I_A = s^2 J(3) = 5.57 s^2 J_{CL}$, which tends to zero as $s \to 0$. *This example is, therefore, also consistent with the surprising result of Subsec. III C, namely, a line charge with a vanishingly small cross section carries a vanishingly small amount of current.* Yet, it is these very small emitting patches that carry the bulk of the anode current, as if the entire cathode surface were emitting, as shown in Fig. 5 of Jassem, and summarized in point (f) in Sec. I. Equation (4.7) offers a resolution to this paradox.

## V. CONCLUDING REMARKS

This paper considers the maximum anode current from an isolated emitting patch with a vanishingly small area, using a new integral equation approach. This idealization bypasses the difficult question concerning the surface electric field distribution on a small emitting patch of finite size. We assume a constant initial velocity for the emitted electrons so that all electrons are moving with a single forward velocity anywhere within the diode. This situation is markedly different from the more realistic model of thermal emission,[3,13,14,33] where a large fraction of emitted electrons may be reflected by the virtual cathode, and only a small fraction of the energetic electrons contributes to the anode current. Both the thermal model for a finite size emission patch, as well as the present model for monoenergetic emission with a vanishingly small dimension, require careful resolution of the potential minimum at the virtual cathode. It is this potential minimum, especially when it is very close to the cathode surface, that causes the most challenge in the numerical solution to both models. This makes a comparison of the present theory difficult because our theory is independent of materials property, whereas thermionic emission depends sensitively on cathode temperature and work function. Despite such limitations, the present theory is in qualitative agreement with the theory and simulations on realistic thermionic cathodes with nonuniform emission.

Prior work on emission from a 2D stripe shows the scaling law, Eq. (1.2), to be valid up to a fairly large value of $d/W = 10$. This scaling law is insensitive to the initial electron velocity assumed, as long as $\Delta \ll 1$. This scaling is not valid in the limit $d/W$ approaching infinity, and it is the main purpose of this paper to analyze this limit ($W \to 0$). Despite the detailed analysis given in this paper, how to continuously generalize the scaling law (1.2) beyond $d/W = 10$ remains an open question. For one thing, it depends on $\Delta$, as shown in Fig. 6. In an extension to thermal emission in the $W \to 0$ limit (which is yet to be done), the critical current would then depend on the work function and on the surface temperature. It follows that the anode current from local "hot spots," which contribute significantly to the total anode current,[14] would depend on the physical causes of strong electron emission at such hot spots.

Nonetheless, the total current reaching the anode is still roughly governed by the 1D CLL, as if the entire cathode were emitting. Once more, this statement is independent of the materials properties, emission mechanism, surface roughness, with hot spots or not, in a 2D or 3D analysis. This remarkable feature is perhaps another aspect of the restriction on the total charge, Q ~ CV, imposed on a diode of vacuum capacitance C, according to an interpretation of CLL in terms of the vacuum gap capacitance.[28,34] Interestingly, if a transverse magnetic field B, in the y-direction in Fig. 1(a), is imposed, the anode current changes only from 100% to 80% of $J_{CL}$ when B is increased from 0% to 90% of the Hull cutoff magnetic field $B_H$, for both zero electron emission velocity[35] and thermal emission model[36] (both assuming uniform emission on the cathode surface, and the z-component of the external magnetic field removed.) These new insights provide some physical basis for the customary use of the 1D CLL to assess the runaway current during diode closure (both nonmagnetized and not fully magnetically insulated ones), by using the instantaneous gap spacing $(d - ut)$ in the 1D CLL, where $u$ is the diode closure velocity.[37]

In writing Eq. (4.4), we assume that the average anode current does not exceed the 1D CL value. There is no reason why this must be so, even though previous analyses[13,14] show that the 1D CL is always approached at a sufficiently high temperature. It is also not known if such a purely static theory may yield a steady state that is reached *nonlinearly* in an emission model that includes a *small thermal* effect,[38] with an injection current density exceeding the 1D Child–Langmuir law.

Finally we note that by finding non-zero solutions for delta-function current density profiles, we have demonstrated that it is possible in principle to exceed $J_{CL}$ locally by an arbitrary large (i.e., "infinite") factor. An interesting question that naturally arises is whether, by spacing such thin sheets periodically by some period $p$, the period-average current density could exceed $J_{CL}$. Our preliminary answer to this question is no, based on initial results from an on-going study.

## ACKNOWLEDGMENTS

This work was supported in part by the Air Force Office of Scientific Research (AFOSR) under Grant Nos. FA9550-20-1-0409 and FA9550-21-1-0184.

## AUTHOR DECLARATIONS

### Conflict of Interest

The authors have no conflicts to disclose.

### Author Contributions

**Y. Y. Lau:** Conceptualization (lead); Formal analysis (equal); Funding acquisition (lead); Investigation (equal); Project administration (lead); Supervision (lead); Writing – original draft (equal); Writing – review & editing (equal). **Dion Li:** Conceptualization (equal); Formal analysis (equal); Investigation (equal); Methodology (equal); Validation (lead); Writing – original draft (equal); Writing – review & editing (equal). **David Chernin:** Conceptualization (equal); Investigation (equal); Validation (equal); Writing – original draft (equal); Writing – review & editing (equal).

## DATA AVAILABILITY

The data that support the findings of this study are available from the corresponding author upon reasonable request.

## APPENDIX A: NUMERICAL SOLUTIONS TO EQ. (3.4)

We numerically solved Eq. (3.4) using the iterative scheme, Eq. (2.10), and the trapezoidal method of integration. To start the iteration ($k = 0$), we set $\overline{\phi}^{(0)}(\overline{z}) = \overline{z}$. The algorithm would halt when $\overline{\phi}^{(k)}(\overline{z}) < -\Delta$ at any value of $\overline{z}$ between (0,1). We remark here that the evaluation of Eq. (3.4) is much more straightforward than that of Eq. (3.10), which becomes apparent in Appendix B, as the integral in Eq. (3.4) is always finite.

As discussed in the paragraph following Eq. (3.5), we ran numerical tests with normalized injection energies of $\Delta = 0$, $10^{-3}$, and 2. We first spot-checked that for values of $J(1)$ below the Jaffe value, Eq. (3.6), the iteration scheme converged after some $k$-steps. We next narrowed down the limiting current values by setting the current density slightly above the values given by Eq. (3.6) and noting the current densities when $\overline{\phi}^{(k)}(\overline{z}) < -\Delta$ for some value of $\overline{z}$. When performing these numerical tests, we found that a larger number of grid points was needed for the potential to converge as the current densities approached the limiting values. Using this procedure, we were able to narrow down the limiting currents with numerical errors of less than 0.5% of the Jaffe value for $\Delta = 0$, $10^{-3}$, and 2 (see Fig. 4).

Thus, in general, we can only find the critical value of $K$ to be within a certain range in the integral equation formulation, for a given $\Delta$. At the lower bound of this range, the iteration (2.10) converges. At the upper bound of this range, the solution $\overline{\phi}(\overline{z})$ does not exist because $\overline{\phi}^{(k)}(\overline{z}) < -\Delta$ for some value of $\overline{z}$ after some $k$-iterations. This approach appears sound because it is able to recover Jaffe's critical $K$ to within 0.5%, with a potential profile showing $\overline{\phi}(\overline{z}) > -\Delta$ for $\Delta = 2$ (Fig. 5), meaning that all electrons move with a forward velocity at the critical $K$, even at the potential minimum as predicted by Jaffe.[32]

## APPENDIX B: NUMERICAL SOLUTIONS TO EQ. (3.10)

Equation (3.10) is also numerically solved using an iterative scheme, Eq. (2.10), and the trapezoidal method of integration. We again set $\overline{\phi}^{(0)}(\overline{z}) = \overline{z}$ to initiate the iteration. In this case, however, it becomes important as to how the singularity, at $\overline{z}_c = \overline{z}$, in the integrand of Eq. (3.10) is treated. Here, we employ two different methods.

The first approach, which we refer to as the "midpoint method," considers a grid of evenly spaced $\overline{z}$ values inclusively between (0,1) and sets the $\overline{z}_c$ grid as the set of midpoints between each $\overline{z}$ value (the $\overline{z}_c$ grid therefore consists of one less point than the $\overline{z}$ grid). This way, the singularity ($\overline{z}_c = \overline{z}$) in the integrand of Eq. (3.10) is "skipped over" and any numerical issues involving this singularity are thus ignored. This, however, also means that the midpoint method becomes less accurate for larger grid spacings as contributions due to this singularity are more significant comparatively.

The second approach, which we call the "singularity inclusion method," considers the two grid cells surrounding the singularity at $\overline{z}_c = \overline{z}$. These two grid cells include the two intervals, $\overline{z} - \Delta\overline{z}_c \le \overline{z}_c \le \overline{z}$, and $\overline{z} \le \overline{z}_c \le \overline{z} + \Delta\overline{z}_c$ where the grid spacing is $\Delta\overline{z}_c$. We divide this region ($\overline{z} - \Delta\overline{z}_c \le \overline{z}_c \le \overline{z} + \Delta\overline{z}_c$) into 20 equally-spaced components. The 18 components away from the singularity may be routinely evaluated, upon linearly interpolating $\overline{\phi}^{(k)}(\overline{z}_c)$ using $\overline{\phi}^{(k-1)}(\overline{z}_c)$ at the two neighboring sub-grid points among the 18 components. In other words, we have, for each of these 18 components,

$$\begin{aligned}\overline{\phi}^{(k)}(\overline{z}) &\simeq \overline{\phi}_a^{(k)}(\overline{z}_c)\\ &= \overline{\phi}^{(k-1)}(\overline{z} + i\Delta\overline{z}_c/N_s)\\ &\quad + \frac{\overline{\phi}^{(k-1)}(\overline{z} + (i+1)\Delta\overline{z}_c/N_s) - \overline{\phi}^{(k-1)}(\overline{z} + i\Delta\overline{z}_c/N_s)}{\Delta\overline{z}_c/N_s}\\ &\quad \times (\overline{z}_c - (\overline{z} + i\Delta\overline{z}_c/N_s)),\\ &\quad \overline{z} + i\Delta\overline{z}_c/N_s < \overline{z}_c < \overline{z} + (i+1)\Delta\overline{z}_c/N_s,\\ &\quad i = 1, 2, \ldots, N_s - 1,\end{aligned} \tag{B1a}$$

$$\begin{aligned}\overline{\phi}^{(k)}(\overline{z}) &\simeq \overline{\phi}_a^{(k)}(\overline{z}_c)\\ &= \overline{\phi}^{(k-1)}(\overline{z} - (i+1)\Delta\overline{z}_c/N_s)\\ &\quad + \frac{\overline{\phi}^{(k-1)}(\overline{z} - i\Delta\overline{z}_c/N_s) - \overline{\phi}^{(k-1)}(\overline{z} - (i+1)\Delta\overline{z}_c/N_s)}{\Delta\overline{z}_c/N_s}\\ &\quad \times \left(\overline{z}_c - (\overline{z} - (i+1)\Delta\overline{z}_c/N_s)\right),\\ &\quad \overline{z} - (i+1)\Delta\overline{z}_c/N_s < \overline{z}_c < \overline{z} - i\Delta\overline{z}_c/N_s,\\ &\quad i = 1, 2, \ldots, N_s - 1,\end{aligned} \tag{B1b}$$

where $N_s = 10$ is the number of components between $\overline{z}_c = \overline{z}$ and $\overline{z}_c = \overline{z} + \Delta\overline{z}_c$ as well as between $\overline{z}_c = \overline{z} - \Delta\overline{z}_c$ and $\overline{z}_c = \overline{z}$. We may now concentrate on the singularity contribution in the region ($\overline{z} - \Delta\overline{z}_c/N_s \le \overline{z}_c \le \overline{z} + \Delta\overline{z}_c/N_s$) by first re-writing Eq (3.10) as

$$\begin{aligned}\overline{\phi}^{(k)}(\overline{z}) = \overline{z} + K_2\int_0^1 &\frac{d\overline{z}_c}{\left(\overline{\phi}^{(k-1)}(\overline{z}_c) + \Delta\right)^{1/2}}\\ &\times \ln\left[\left|\sin\left(\frac{\pi}{2}(-\overline{z} + \overline{z}_c)\right)\right|\right]\\ - K_2\int_0^1 &\frac{d\overline{z}_c}{\left(\overline{\phi}^{(k-1)}(\overline{z}_c) + \Delta\right)^{1/2}}\ln\left[\sin\left(\frac{\pi}{2}(\overline{z} + \overline{z}_c)\right)\right],\\ 0 \le \overline{z} \le 1, &\end{aligned} \tag{B2}$$

where the third term on the RHS of Eq. (B2) may be computed accurately as the logarithmic term never diverges. Upon approximating $\ln\left[|\sin(\pi(-\overline{z} + \overline{z}_c)/2)|\right]$ as $\ln[|-\overline{z} + \overline{z}_c|] + \ln[\pi/2]$ when $\overline{z}_c$ is very close to $\overline{z}$, the contribution due to the second term on the RHS of Eq. (B2) from the immediate vicinity of the singularity ($\overline{z} - \Delta\overline{z}_c/N_s \le \overline{z}_c \le \overline{z} + \Delta\overline{z}_c/N_s$) may be computed in closed form.

For each $(K_2, \Delta)$ pair, we used the midpoint and singularity inclusion methods to validate each other over varying $N$ ($N$ = total number of grid points over the interval, $0 < \overline{z} \le 1$). Figure 7 shows an example of this validation. We were able to obtain accurate estimates of the 2D limiting current $K_2$ by plotting the minimum value of $(\overline{\phi}(\overline{z}) + \Delta)$ vs $N$, finding the value to which $\min(\overline{\phi}(\overline{z}) + \Delta)$ approaches at large $N$, and then adjusting $K_2$ such that $\min(\overline{\phi}(\overline{z}) + \Delta)$ approaches zero. We have empirically determined that the midpoint method tends to overestimate the exact value of $\min(\overline{\phi}(\overline{z}) + \Delta)$ for smaller $N$, as shown in Fig. 7, and that the singularity inclusion method tends to oscillate about $\min(\overline{\phi}(\overline{z}) + \Delta)$ with decreasing amplitude as $N$ increases.

We have additionally tried a "polynomial method" of numerically evaluating Eq. (3.10) where we assume that $\overline{\phi}(\overline{z})$ takes on the form, for a general $\Delta$,

$$\overline{\phi}^{(k)}(\overline{z}) \simeq \sum_{m=1}^{M} a_m \overline{z}^m, \tag{B3}$$

in the argument inside the square-root of the integrand of Eq. (3.10). We may then solve for $\overline{\phi}^{(k)}(\overline{z})$ by determining the coefficients $a_1, \ldots, a_M$. We notice, however, that our results only converge when $M \gg 10$, making this method quite computationally expensive as it involves the inversion of $M \times M$ matrices.

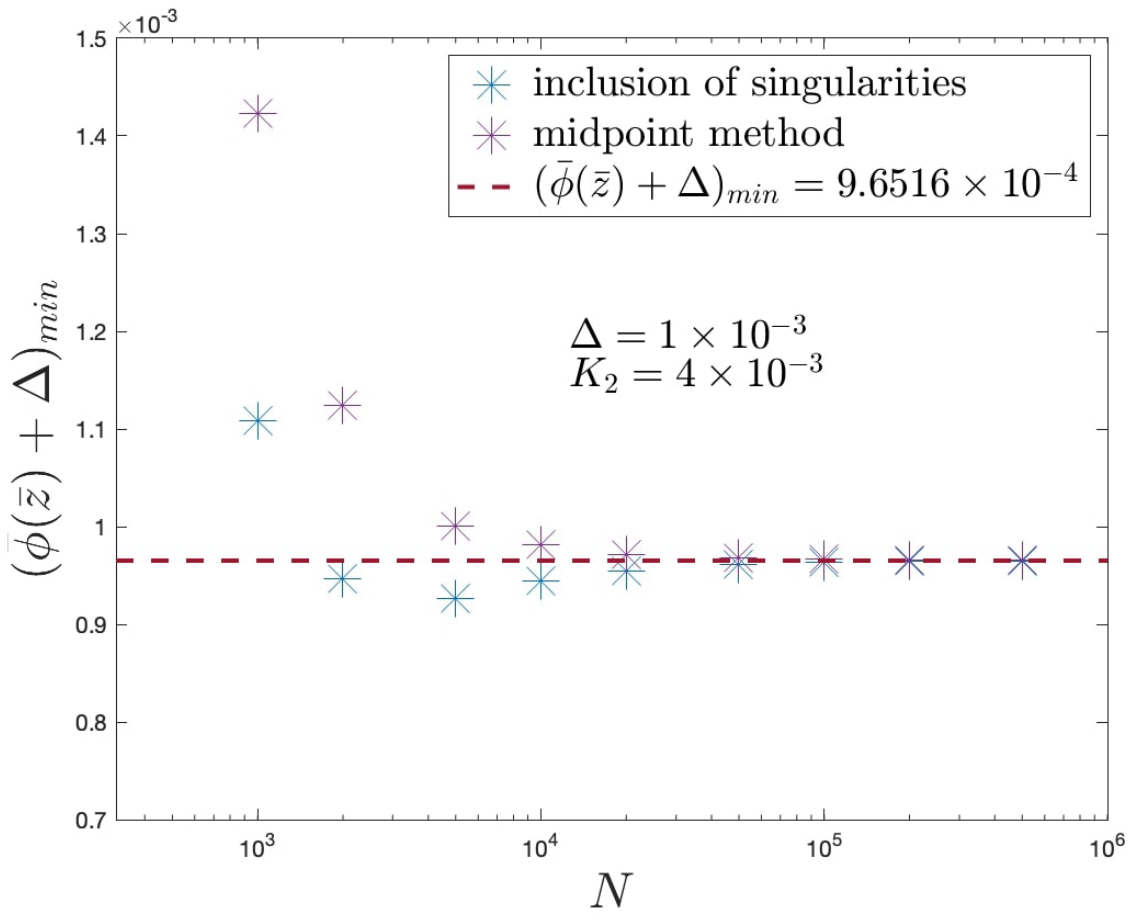

**FIG. 7.** Convergence of the value of $\min(\overline{\phi}(\overline{z})+\Delta)$ as function of the number of grid points $N$ for $\Delta = 1\times 10^{-3}$ and $K_2 = 4\times 10^{-3}$ using the two different numerical algorithms described in Appendix B.

## APPENDIX C: PROOF OF NULL SOLUTION TO EQ. (3.10) WHEN $\Delta = 0$

Equation (3.10) may be rewritten as with $\Delta = 0$,

$$\overline{\phi}(\overline{z}, K_2) = \overline{z} + K_2 S(\overline{z}, K_2), \quad 0 \le \overline{z} \le 1, \tag{C1a}$$

$$S(\overline{z}, K_2) = \int_0^1 \frac{d\overline{z}_c}{\left(\overline{\phi}(\overline{z}_c, K_2)\right)^{1/2}} \ln\left[\frac{\left|\sin\left(\frac{\pi}{2}(-\overline{z}+\overline{z}_c)\right)\right|}{\sin\left(\frac{\pi}{2}(\overline{z}+\overline{z}_c)\right)}\right]. \tag{C1b}$$

We note that $\overline{\phi}(\overline{z}) \le \overline{z}$ since $S(\overline{z}, K_2) \le 0$. It thus follows that

$$S(\overline{z}, K_2) \le \int_0^1 \frac{d\overline{z}_c}{\overline{z}_c^{1/2}} \ln\left[\frac{\left|\sin\left(\frac{\pi}{2}(-\overline{z}+\overline{z}_c)\right)\right|}{\sin\left(\frac{\pi}{2}(\overline{z}+\overline{z}_c)\right)}\right] \equiv S^{(0)}(\overline{z}), \tag{C2}$$

where $S^{(0)}(\overline{z})$ refers to $S(\overline{z})$ in the zeroth iteration ($k = 0$) since we take $\overline{\phi}^{(0)}(\overline{z}) = \overline{z}$. Next, we concentrate on small $\overline{z}$, in particular $\overline{z} \to 0^+$, as we wish to prove nonexistence of solution even for a very small, nonzero value of $K_2$. We write

$$S^{(0)}(\overline{z}) = S_1(\overline{z}) + S_2(\overline{z}), \tag{C3a}$$

$$S_1(\overline{z}) = \int_0^{\overline{z}} \frac{d\overline{z}_c}{\overline{z}_c^{1/2}} \ln\left[\frac{\left|\sin\left(\frac{\pi}{2}(-\overline{z}+\overline{z}_c)\right)\right|}{\sin\left(\frac{\pi}{2}(\overline{z}+\overline{z}_c)\right)}\right]$$
$$\simeq \int_0^{\overline{z}} \frac{d\overline{z}_c}{\overline{z}_c^{1/2}} \ln\left[\frac{\overline{z}-\overline{z}_c}{\overline{z}+\overline{z}_c}\right] = \overline{z}^{1/2} \int_0^1 \frac{d\xi}{\xi} \ln\left[\frac{1-\xi}{1+\xi}\right], \tag{C3b}$$

$$S_2(\overline{z}) = \int_{\overline{z}}^1 \frac{d\overline{z}_c}{\overline{z}_c^{1/2}} \ln\left[\frac{\left|\sin\left(\frac{\pi}{2}(-\overline{z}+\overline{z}_c)\right)\right|}{\sin\left(\frac{\pi}{2}(\overline{z}+\overline{z}_c)\right)}\right] \equiv S_{2A}(\overline{z}) + S_{2B}(\overline{z}), \tag{C3c}$$

$$S_{2A}(\overline{z}) = \int_{\overline{z}}^{\epsilon} \frac{d\overline{z}_c}{\overline{z}_c^{1/2}} \ln\left[\frac{\left|\sin\left(\frac{\pi}{2}(-\overline{z}+\overline{z}_c)\right)\right|}{\sin\left(\frac{\pi}{2}(\overline{z}+\overline{z}_c)\right)}\right], \tag{C3d}$$

$$S_{2B}(\overline{z}) = \int_{\epsilon}^1 \frac{d\overline{z}_c}{\overline{z}_c^{1/2}} \ln\left[\frac{\left|\sin\left(\frac{\pi}{2}(-\overline{z}+\overline{z}_c)\right)\right|}{\sin\left(\frac{\pi}{2}(\overline{z}+\overline{z}_c)\right)}\right], \tag{C3e}$$

where $\xi = \overline{z}_c/\overline{z}$ in Eq. (C3b) and we have chosen an $\epsilon$ such that $\overline{z} \ll \epsilon \ll 1$ (for $\overline{z} \to 0^+$). This allows us to approximate

$$S_{2A}(\overline{z}) \simeq \int_{\overline{z}}^{\epsilon} \frac{d\overline{z}_c}{\overline{z}_c^{1/2}} \ln\left[\frac{-\overline{z}+\overline{z}_c}{\overline{z}+\overline{z}_c}\right] = \overline{z}^{1/2} \int_1^{\epsilon/\overline{z}} \frac{d\xi}{\xi^{1/2}} \ln\left[\frac{\xi-1}{\xi+1}\right]$$
$$\simeq \overline{z}^{1/2} \int_1^{\infty} \frac{dx}{x^{1/2}} \ln\left[\frac{x-1}{x+1}\right] = \overline{z}^{1/2} \int_0^1 \frac{d\xi}{\xi^{3/2}} \ln\left[\frac{1-\xi}{1+\xi}\right], \tag{C4}$$

where $\xi = 1/x$ in the last integral, which is a finite constant.

We shall momentarily show that $S_{2B}(\overline{z}) \ll S_{2A}(\overline{z})$ in magnitude, whence Eqs. (C3) and (C4) give

$$S^{(0)}(\overline{z}) = S_1(\overline{z}) + S_{2A}(\overline{z}) + S_{2B}(\overline{z}) \simeq S_1(\overline{z}) + S_{2A}(\overline{z})$$
$$\simeq \overline{z}^{1/2} \int_0^1 \frac{d\xi}{\xi}\left(1 + \frac{1}{\xi^{1/2}}\right) \ln\left(\frac{1-\xi}{1+\xi}\right) = -2\pi \overline{z}^{1/2}. \tag{C5}$$

To show that $S_{2B}(\overline{z})$ is small compared with $S_{2A}(\overline{z})$ in magnitude, we expand $\sin(\pi(\overline{z} \pm \overline{z}_c)/2)$ for vanishingly small $\overline{z}$, recalling that $\overline{z} \ll \epsilon \ll 1$ and $\epsilon < \overline{z}_c < 1$ in Eq. (C3e). We then obtain from Eq. (C3e),

$$S_{2B}(\overline{z}) \simeq \int_{\epsilon}^1 \frac{d\overline{z}_c}{\overline{z}_c^{1/2}} \ln\left[1 - \pi\overline{z}\frac{\cos\left(\frac{\pi\overline{z}_c}{2}\right)}{\sin\left(\frac{\pi\overline{z}_c}{2}\right)}\right]$$
$$\simeq -\pi\overline{z}\int_{\epsilon}^1 \frac{d\overline{z}_c}{\overline{z}_c^{1/2}} \frac{\cos\left(\frac{\pi\overline{z}_c}{2}\right)}{\sin\left(\frac{\pi\overline{z}_c}{2}\right)}$$
$$\sim -\pi\overline{z}\frac{4}{\pi\epsilon^{1/2}} = -4\overline{z}^{1/2}(\overline{z}/\epsilon)^{1/2} \ll \left|S_{2A}(\overline{z})\right|. \tag{C6}$$

In the second integral of Eq. (C6), we estimate its value by noting that the dominant contribution comes from its lower limit ($\overline{z}_c \sim \epsilon$). Therefore, for very small values of $\overline{z}$, we obtain from Eqs. (C1a) and (C2),

$$\overline{\phi}^{(1)}(\overline{z}, K_2) \le \overline{z} + K_2 S^{(0)}(\overline{z}). \tag{C7}$$

Substitution of Eq. (C5) into Eq. (C7) yields $\overline{\phi}^{(1)}(\overline{z}, K_2) \le 0$ when $\overline{z}^{1/2} \le 2\pi K_2$. This means that Eq. (C1a) has no iterative, real solution of $\overline{\phi}$ regardless of how small is $K_2$, as long as $K_2$ is nonzero. This completes the proof.